\documentclass[twocolumn,amsmath,amssymb]{revtex4}
\usepackage[colorlinks=true, pdfstartview=FitV, linkcolor=magenta, citecolor=red, urlcolor=blue]{hyperref}
\usepackage{epsfig,color}
\usepackage{xcolor}
\usepackage{epstopdf}
\usepackage{amssymb}
\usepackage{verbatim}
\usepackage{multirow}
\usepackage{latexsym}
\usepackage{epsfig}
\usepackage{epstopdf}
\usepackage{graphicx}
\usepackage{amssymb}
\usepackage{amsmath}
\usepackage{dcolumn}
\usepackage{bm}
\usepackage{color}
\usepackage{comment}
\begin{document}

\title{Differential topology and micro-structure of black hole in Einstein-Euler-Heisenberg spacetimes with exponential entropy}

\author{Muhammad Yasir$^1$}
\email{yasirciitsahiwal@gmail.com (corresponding author)}
\affiliation{$^1$Institute of Fundamental Physics and Quantum Technology, Department of Physics, School of Physical Science and Technology, Ningbo University, Ningbo, Zhejiang 315211, People's Republic of China}
\author{Tong Lining$^{2*}$}
 \email{tongln@shu.edu.cn (corresponding author)}
 \affiliation{$^{2}$Department of Mathematics, Shanghai University  and Newtouch Center for Mathematics of Shanghai University,  Shanghai, 200444, People's Republic of China}
\author{Kazuharu Bamba$^3$}
 \email{bamba@sss.fukushima-u.ac.jp}
 \affiliation{$^3$Faculty of Symbiotic Systems Science\\
   Fukushima University, Fukushima 960-1296, Japan}
\begin{abstract}
Exact black holes in the Einstein Euler-Heisenberg theory are explored with an exponential entropy framework by using the topological current $\Psi$-mapping theory. The topology classes are investigated through the canonical, mixed, and grand canonical ensembles. In particular, the magnetic charge is fixed for the canonical ensemble, whereas the magnetic potential is included for the mixed ensemble and the grand canonical ensemble with maintaining its consistency through the magnetic potential. The topological charges are analyzed for each ensemble through critical points. As a result, it is found that the canonical, mixed, and grand canonical ensembles lead to either $1$, $-1$, or no generation/annihilation points. Moreover, it is shown how temperature and heat capacity depend on the horizon radius in order to verify the stability of a black hole. Furthermore, the behavior of the thermodynamic curvatures of a black hole is investigated through the geometric methods. \\
\textbf{Keywords}: Black hole, Einstein Euler-Heisenberg theory; Topology classes; Exponential entropy, Micro-structures.
\end{abstract}
\maketitle

\section{Introduction}

One of the most intriguing and challenging areas of study in general relativity (GR) and modified theories of gravity is the thermodynamic structure of black hole (BH) \cite{1}. The charged BH is an interesting approach that can provide the valuable insights into the fundamental characteristics of the modified theory of gravities.
It would be fascinating to study the thermodynamic topology of charged BHs, where the interplay between circular motion and  electromagnetic forces can sustain equilibrium. According to Bekenstein,  a stationary classical solution does not exist however Higgs field has a homogenous configuration and takes the value based on the potential slope. It is not equivalent to any stationary quantum state \cite{3}. The classical stationary stable solution corresponds to a field configuration that minimizes the potential. In this case, the quantum theory has small perturbations away from it and this regarded as excitations of the field above the minimal state, which classically fluctuates around it.
 
The Einstein–Euler–Heisenberg theory is developed to study how electromagnetic fields behave differently in extremely strong gravitational or electromagnetic environments. It incorporates nonlinear effects predicted by quantum field theory and string theory, which become significant in such high-energy conditions \cite{Ruffini:2013hia}. In this scenario, this system expands on classical electromagnetism and allows the investigation of phenomena like photon scattering and vacuum birefringence. Therefore, many features have been discussed in BH physics, where the nonlinear effects influence thermodynamic properties and quasinormal modes (QNMs) \cite{4L, Breton:2021mju}. Moreover, the characteristics of electromagnetic fields near BHs have been extensively investigated, providing key perspectives into quantum-gravitational dynamics \cite{Karakasis:2022xzm}.

On the other hand, a classical stationary but unstable solution is a configuration where the field has maximum potential. Tachyonic excitations are the new interpretation of such perturbations in quantum theory.
However, a classical stationary but unstable solution reflects to a field configuration with maximum potential energy. 
In quantum theory, such perturbations are reformulated as tachyonic excitations. The heat quantity and state parameters are analyzed according to the four well-known laws of BH mechanics \cite{3}. The Hawking-Page phase transition \cite{3a} establishes a fundamental connection between thermodynamics and  gravity, marking a significant development in BH physics.

Numerous methods are employed to investigate the thermodynamic properties of topology in extended phase space, where the cosmological constant represents a thermodynamic pressure. This study classifies into three categories based on topological numbers such as $+1$, $0$, or $-1$. A novel advancement in BH thermodynamic topology is a $\Psi$-mapping theory. This framework, referred to as $\Psi$-mapping process, is applied within the thermodynamic space of a BH to study the distinctive features of this innovative technique, originally introduced in  Refs. \cite{i1,1.4,1.5}. 

The BH temperature $T$ can be computed as a function of its entropy $S$, state parameter $P$, and the other thermodynamic factors through relation as $T=T\left(S, P, x^i\right)$. Here, thermodynamic parameters are represented by $x^i$. In this case, the two-dimensional vectors in  $\Psi$ are known as $\Psi ^\theta=\left(\partial_\theta\Psi\right)_{S, x^i}$,  \mbox{and} \quad $\Psi^S=\left(\partial_S \Psi\right)_{\theta, x^i}$, respectively. These vectors are defined in Duan's current $\Psi$-mapping theory \cite{1.4,1.5} as suits with the factor $1/\sin \theta$. Therefore, the existence of $\theta$ in $\Psi$ provides the zero point of the vector field $\Psi$ at $\theta=\pi/2$. This technique can be used to estimate the critical points, and the topological charge $j^\mu$ likewise satisfies the conservation law follows as $\Psi^a\left(x^i\right)=0$.

The topology in thermodynamics has been discussed in extended phase space of various BHs in  Refs. \cite{i4, Wei:2022dzw,i5,i6,i8,i8A,i8B, i9}, where $w_i$, $j^0$ and $\beta_i$ represent the winding number of $i$-th zero points, the density of the topological current and the Hope index, respectively. The total topological number of BH is computed by adding the individual winding numbers related with each of its charges.
These considerations have been investigated as topological defects of numerous BHs in Refs. \cite{i12,Bhattacharya:2024bjp,i13,i14,i15,i21,i21A}. 

Several attempts have been made to discuss a BH's thermodynamic behaviors by using the Riemannian geometry.  In particular, the Weinhold and Ruppeiner metrics, formulated as the Hessian matrix of internal energy and entropy, respectively; this used to obtain the direct results between phase transitions and curvature singularities \cite{g3,g4,g8}. 
 As a result, we study a novel version of Ruppeiner's metric \cite{g3}, which uses Legendre transformations to relate the thermodynamic potentials to the mass rather than the entropy while utilizing specific heat to measure the stability of BH. By using the Ruppeiner, Weinhold and Hendi–Panahiyan–Eslam Panah–Momennia (HPEM) formulas, we also examine the thermodynamic curvatures behavior of BH \cite{g9, Aman:2003ug}. For certain constants, it is analyzed that thermodynamic curvatures display the BH's repulsive and attractive characteristics. This demonstrated a one-to-one correlation between thermal energy divergences and thermal quantities curvatures in this novel type of thermodynamic geometry \cite{Aman:2003ug}. In contrast, the Riemannian thermodynamic structure of the Geometrothermodynamics (GTD) approach which adheres to Legendre invariance. Indeed, the impetuous application of the modified and natural thermodynamic variables in BH thermodynamics may highlight certain inconsistencies in GTD \cite{g15,g16,g19}. 

In this paper, we discuss how ingoing and outgoing flows correspond to stable and unstable thermodynamic behavior, respectively. The topological charge, obtained from the BH’s free energy, acts as a stability index. In Einstein-Euler-Heisenberg BHs, the exponential form of the entropy defines this free energy landscape. A topological charge of $+1$, typically labels a stable BH region with positive heat capacity. However, charge of $-1$ provides an unstable region.
In addition, we examine the geometric structure of the Ruppiner, Weinhold, HPEM and GTD formalisms to use an exponential entropy framework to discuss the phase transition of BH in Einstein Euler-Heisenberg theory. By employing geometric methods, we derive detailed insights into the statistical properties and thermodynamic interactions of BHs.
Further, by employing geometric methods, we derive detailed insights into the statistical properties and thermodynamic interactions of BH in instein-Euler-Heisenberg theory. 
This approach uncovers novel microstructure signatures such as phase transitions and critical phenomena that elucidate fundamental geometric and quantum features of BHs.

The organization of the paper is as follows. 
In Sec.~\ref{Sec:A}, we study a short review of exact BH in Einstein Euler-Heisenberg theory.
In Sec.~\ref{Sec:B}, we study the BH  thermodynamic topology through the canonical ensemble.
In Sec.~\ref{Sec:C}, in the canonical ensemble, we introduce topological defects.
In Sec.~\ref{Sec:D}, our focus is on BH in Einstein Euler-Heisenberg theory as a mixed ensemble.
In Sec.~\ref{Sec:E}, we study topological defects of considered BH solution through mixed ensemble.
In Sec.~\ref{Sec:F}, we discuss the thermal geometries with an exponential framework.
In Sec.~\ref{Sec:G}, we study the emission energy with an exponential framework.
Finally, we present conclusions in Sec.~\ref{Sec:L}. 

\section{Black hole spacetimes in Einstein Euler-Heisenberg theory} \label{Sec:A}

The Einstein-frame action admits us to employ a simplified model, which takes the form \cite{BM1,BM4}
\begin{eqnarray}\label{Eq1}
\nonumber
 \mathcal{S}&=&\frac{1}{16 \pi} \int \mathrm{d}^4 x \sqrt{-g}[\mathcal{R}-e^{-2 \phi} \mathcal{F}^2-2 \nabla^\mu \phi \nabla_\mu \phi\\  \label{b1}
 &-&f(\phi)(2 \alpha \mathcal{F}_\beta^\alpha \mathcal{F}_\gamma^\beta \mathcal{F}_\delta^\gamma \mathcal{F}_\alpha^\delta-\beta \mathcal{F}^4)], 
\end{eqnarray}
where the Ricci scalar is denoted by $\mathcal{R}$, the usual Faraday scalars are represented by $\mathcal{F}^2 \equiv \mathcal{F}_{\mu \nu} \mathcal{F}^{\mu \nu} \sim$ $\mathbf{E}^2-\mathbf{B}^2$ and $\mathcal{F}^4 \equiv \mathcal{F}_{\mu \nu} \mathcal{F}^{\mu \nu} \mathcal{F}_{\alpha \beta} \mathcal{F}^{\alpha \beta}$, 
$\mathcal{F}_{\mu \nu} = \partial_\mu \mathcal{A}_\nu-\partial_\nu \mathcal{A}_\mu$ represents the usual field strength tensor and $\mathcal{A}_\nu$ is the U(1) gauge field.  Moreover, the coefficients of $\alpha$ and $\beta$ are coupling constants with dimensions of (length)$^2$. 
Furthermore, the scalar field $\phi$ corresponds to the dilaton coupled to 
the electromagnetic fields through the interrelated scalar coupling function $f(\phi)$. Here, both $\phi$ and $f(\phi)$ are dimensionless amounts. The inclusion of the Gauss-Bonnet invariant via non-diagonal reduction gives a conception to this kind of the field-theoretic gravitational action in $\left(\ref{Eq1}\right)$.  This theory admits the Gibbons-Maeda-Garfinkle-Horowitz-Strominger (GMGHS) BH \cite{BM1,BM5} and it leads to an accurate solution for setting $f(\phi)=0$. From the action in $\left(\ref{Eq1}\right)$, 
the field equations read \cite{BM1}
\begin{eqnarray} \nonumber
G_{\mu \nu}&=&2 \partial_\mu \phi \partial_\nu \phi-g_{\mu \nu} \partial^\alpha \phi \partial_\alpha \phi+2 e^{-2 \phi}\bigg(\mathcal{F}_\mu^\alpha \mathcal{F}_{\nu \alpha}\\ \nonumber
&-&\frac{1}{4} g_{\mu \nu} \mathcal{F}^2\bigg)+f(\phi)\bigg\{8 \alpha \mathcal{F}_\mu^\alpha \mathcal{F}_\nu^\beta\mathcal{F}_\alpha^\eta \mathcal{F}_{\beta \eta}-\alpha g_{\mu \nu}\\  \nonumber
&&\mathcal{F}_\beta^\alpha\mathcal{F}_\gamma^\beta \mathcal{F}_\delta^\gamma \mathcal{F}_\alpha^\delta-4 \beta \mathcal{F}_\mu^{\xi} \mathcal{F}_{\nu \xi} \mathcal{F}^2+\frac{1}{2} g_{\mu \nu} \beta \mathcal{F}^4\bigg\},
\\ 
\label{bb1}
\end{eqnarray}
\begin{eqnarray} \nonumber
4 \square \phi&=&-2 e^{-2 \phi} \mathcal{F}^2+\frac{d f(\phi)}{d \phi}\bigg(2 \alpha \mathcal{F}_\beta^\alpha \mathcal{F}_\gamma^\beta \mathcal{F}_\delta^\gamma \mathcal{F}_\alpha^\delta-\beta \mathcal{F}^4\bigg),
\\ 
\label{bba1}
\end{eqnarray}
and 
\begin{eqnarray}\nonumber
&&\partial_\mu \bigg\{\sqrt{-g}\bigg[4 \mathcal{F}^{\mu \nu}\left(2 \beta f(\phi) \mathcal{F}^2-e^{-2 \phi}\right)\\  \label{b2}
&-&16 \alpha \mathcal{F}^\mu_\kappa\mathcal{F}^\kappa{ }_\lambda \mathcal{F}^{\nu \lambda}\bigg]\bigg\}=0,
\end{eqnarray}
where, $\square$ presented by d'Alembert operator and we are interested in expanding the GMGHS solution \cite{BM5} by accounting for higher-order electromagnetic invariants $\mathcal{F}^4$ and $\mathcal{F}_\beta^\alpha \mathcal{F}_\gamma^\beta \mathcal{F}_\delta^\gamma \mathcal{F}_\alpha^\delta$, respectively. For further mathematical details on these expressions, we refer the reader to \cite{BM1,BM5}.
We introduce the broadest spherically symmetric metric ansatz in the form \cite{BM1}
\begin{eqnarray}\label{b3}
\mathrm{d} s^2=-B(r) \mathrm{d} t^2+\frac{\mathrm{d} r^2}{B(r)}+[R(r)]^2 \mathrm{~d} \Omega^2,
\end{eqnarray}
where $\mathrm{d} \Omega^2=\mathrm{d} \theta^2+\sin ^2 \theta \mathrm{~d} \varphi^2$.  
However, as emphasized earlier and further elaborated below, a dilaton-independent term in $f(\phi)$ is essential for the analytic treatment of the BH solution. In above discussion, such a term can arise from higher-order string loop corrections in the underlying string-theory model. Note that in this scenario, $\varphi$ consistently refers to the azimuthal coordinate, whereas $\phi$ denotes the scalar field. Additionally, we take into account magnetic and electric charges using the following four vectors which are consistent with spherical symmetry as 
\begin{eqnarray}\label{b4}
\mathcal{A}_\mu=\left(V(r), 0,0, \cos (\theta) Q_m\right),
\end{eqnarray}
where the magnetic charge associated with the BH is denoted by $Q_m$. The construction $\varphi$ component of the Maxwell equations to solve this ansatz for the electromagnetic field, it is determined that the scalar field inherits the spacetime symmetries such as $\phi \equiv \phi(r)$. As a result, we find
\begin{eqnarray} \nonumber
2 \alpha \mathcal{F}_\beta^\alpha \mathcal{F}_\gamma^\beta \mathcal{F}_\delta^\gamma \mathcal{F}_\alpha^\delta-\beta
\mathcal{F}^4=\frac{4(\alpha-\beta) Q_m^4}{[R(r)]^8}\\  \label{b5}
+\frac{8 \beta Q_m^2\left[V^{\prime}(r)\right]^2}{[R(r)]^4}+4(\alpha-\beta)\left[V^{\prime}(r)\right]^4.
\end{eqnarray}
In the case of $\alpha=\beta$, this will disappear on the condition that neither electric nor magnetic configurations are considered.  As earlier explained, prime stands for derivative with respect to $r$. Since it is extremely challenging to integrate Maxwell's equation for the dyonic scenario, we will focus on pure magnetic fields, such as $V(r)=0$. Therefore, if $\alpha \neq \beta$ is satisfied, both of these non-linear electrodynamics factors would contribute. This solution have been developed for the scalar free with setting $\phi=0$ and $f(\phi=0)=1$ as \cite{BM1}
\begin{eqnarray}\label{b6}
B(r)=1+\frac{Q_m^2}{r^2}+\frac{2(\alpha-\beta) Q_m^4}{5 r^6}-\frac{2 M}{r},
\end{eqnarray}
and $R(r)=r$. This metric function resembles the Einstein-Euler-Heisenberg BH \cite{BM4}. The continued existence of the contribution depends only on the values of the coupling parameters $\alpha$ and $\beta$. Remember that the higher-order electromagnetic term has no influence at all when $\alpha=\beta$. However, the BH solution with scalar hair in the Euler-Heisenberg theory was investigated in Ref.~\cite{BM4}.
\section{Thermodynamic topology in canonical ensemble}\label{Sec:B}

In this section, we will expand upon the previous discussion to derive the expressions for the topological number as
\cite{i5,i6,i8,i9}
\begin{equation}\label{i1}
Q=\int_{\Sigma} j^0 d^2 x=\sum_{i=1}^N w_i=\sum_{i=1}^N \beta_i n_i.
\end{equation}
The topology in BH thermodynamics has been investigated in Refs.~\cite{i12,i13}. 
Here, $w_i$, $j^0$, and $\beta_i$ are the Hope index, the density of the topological current, and assign a winding number ($i-$th) zero points of $\Psi$ contained in $\Sigma$, respectively. To initiate our study, we first introduce the propagating free energy $F$, defined as
\begin{equation}\label{i2}
F=E-\frac{S_E}{\tau}.
\end{equation}
The free energy is introduced and defined as follows to start the analysis.
The potential energy, entropy, and quantity with a dimension of time are represented by the variables $E$, $S$, and $\tau$, respectively.
Here, $\tau$ is an auxiliary parameter with the dimension of time, interpreted as the inverse temperature of a cavity surrounding the BH. For subsequent analysis, an arbitrary length scale is determined by the size of the cavity enclosing the BH. For sufficiently large $\tau$, such as $\tau=t_1$, there exist one and two intersection points for the Schwarzschild and Kerr BHs, respectively. These points precisely fulfill the condition $\tau=1/T$, and thus correspond to on-shell BH solution with the characteristic temperature $T=1/\tau$. It should be noted that when this parameter coincides with the Hawking temperature, the solution satisfies Einstein’s field equations, thereby rendering the free energy on-shell. The comparative analyses with other topological invariants could elucidate its distinctive contributions. Typically, incorporating $\tau$ within quantum gravitational corrections would establish a more comprehensive and resilient framework for the classification of BH thermodynamics.
We find the following expression for a vector field $\Psi$ in terms of free energy
\begin{equation}\label{i4}
\Psi=\left(\frac{\partial {F}}{\partial r_{h}},-\cot \Theta \csc \Theta\right).
\end{equation}
The vector $\Psi$ has a zero point at $\Theta=\pi/2$. For the mass function, we calculate the function $B(r_{h})$ near the event horizon which is represented as $r=r_{h}$. 
In particular, we emphasize that the exponential entropy formalism remains consistent with the modified geometric thermodynamic and microstructure of Einstein–Euler–Heisenberg gravity, where nonlinear electrodynamics introduces corrections to the standard Bekenstein–Hawking entropy. The exponential framework provides a regularized description of microstates and naturally incorporates higher–order corrections arising from the Einstein–Euler Heisenberg Lagrangian. This compatibility thus offers a more generalized and robust description of BH entropy beyond the classical area law.
The modified entropy expression is given as \cite{H5,C1}
\begin{equation}\label{iA2}
S_E=\frac{\mathcal{A} \ln 2}{8 \pi \gamma l_p^2}+e^{-\mathcal{A} \ln 2 / {(8 \pi \gamma l_p^2})}.
\end{equation}
Thus, for the choice $\gamma=\ln2 / 2 \pi$, the main term Bekenstein-Hawking result is reconstructed, additionally, a correction to the classical finding that is exponentially suppressed is produced.
\begin{figure}
\includegraphics[width=16pc]{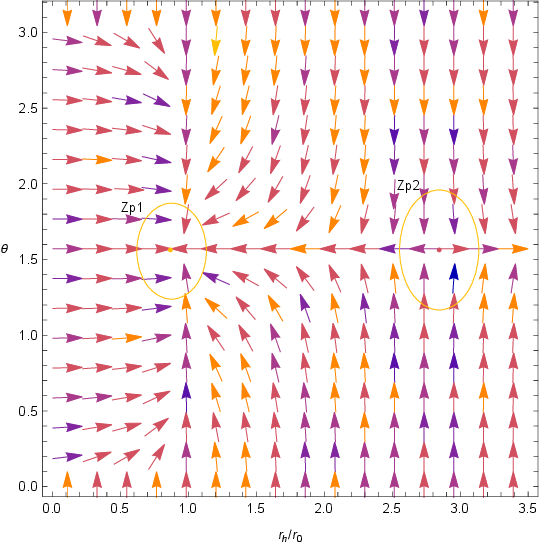}
\caption{\label{fig2} The field of unit vectors with fixed values of $\alpha =1.1$, $Q_m=1$, and $\tau =30$.}
\end{figure}
This formalism takes into account the BH entropy which provides the geometric features of the BH horizon. In order to achieve this,  by incorporating an exponential factor $e^{-\pi r_h^2}$, the entropy formula is modified and its application will expand.
We applied the series expansion of the exponential entropy, given by $S_E \approx e^{-\pi r_h^2} + \pi r_h^2$, from this relation, the horizon radius can be expressed as $r_h = \left(\frac{2(S_E - 1)}{\pi^2}\right)^{1/4}$.
With positive values of the parameters lead to the stability of the system, whereas negative values indicate the instability of BH \cite{C2,C3,C3A,C3B,C4,C5}. 
The mass  parameter of BH can be written as
\begin{widetext}
\begin{eqnarray}  \label{1}
M&=&\frac{\pi^3 {Q_m}^4 (\alpha -\beta)+5({S_E}-1)[\pi  {Q_m}^2 + \sqrt{2} ({S_E}-1)^{1/2}]}{10 \sqrt[4]{2} \sqrt{\pi}({S_E}-1)^{5/4}}.
\end{eqnarray}
\end{widetext}
This modification is essential to study the behavior of  BH in Einstein Euler-Heisenberg theory in thermodynamics and physical effects, where $r_h$  stands for the horizon radius and temperature relation  as $T=\partial M/\partial {S_E}$.
By utilizing the positive and negative signs of the roots which determine the divergence points of the heat capacity and thermal stability of BH are  examined.
The characteristics of the BH horizon region and its entropy linked to the physical content of the exponential term. This non-standard component could affect the stability, quantum characteristics and thermodynamics of BH. The heat capacity relation is provided as
\begin{eqnarray} \nonumber
C&=&\frac{T}{\frac{\partial^2 M}{\partial S_E^2}}\\ \nonumber
&=& \frac{2 {\pi }^{2} r_h^4 [-{{ Q_m}}^{2}{r_h}^{4}+{r_h}^{6}-2 {Q_m}^{4} (\alpha-\beta)^{4}]^{4}}{5{{Q_m}}^{2}{r_h}^{4}-3{r_h}^{6} +18  {Q_m}^4 (\alpha-\beta)^{4}},\\ \label{2a} 
\end{eqnarray}
where 
\begin{eqnarray} \nonumber 
\frac{\partial^2 M}{\partial S_E ^2}
&=&({2}^{3/4} [5\pi {{Q_m}}^{2} (-1+{S_E}) -3\sqrt {2} (-1+{S_E})^{3/2}\\  
&+&9{\pi}^{3}{{Q_m}}^{4} (\alpha-\beta)])/{64  {\sqrt {\pi}} 
(-1+{S_E})^{13/4}}.
\end{eqnarray}
In that case, heat capacity has positive behavior by ensuring the system’s stability. 
Then, we eliminate the state parameter by applying the formula $\left(\partial_S T\right)_{P x^i}=0$, and this formation is examined by a new potential $\Psi$, which is also called Duan's current potential $\Psi=T\left(S, x^i\right) /{\sin \theta}$ as
\begin{widetext}
\begin{eqnarray} \label{2c}
\Psi =\frac{-\pi {Q_m}^2 (e^{-\pi r_h^2}+\pi  r_h^2-1)+\sqrt{2} (e^{-\pi r_h^2}+\pi r_h^2-1)^{3/2}+\pi^3 {Q_m}^4 (\beta -\alpha)}{[8 \sqrt[4]{2} \sqrt{\pi } (e^{-\pi r_h^2}
+\pi  r_h^2-1)]^{9/4} \sin (\theta)},
\end{eqnarray}
\end{widetext}
where temperature parameter can be calculated as 
\begin{widetext}
\begin{eqnarray} \label{2d}
T=\frac{-\pi {Q_m}^2 (e^{-\pi r_h^2}+\pi  r_h^2-1)+\sqrt{2} (e^{-\pi r_h^2}+ \pi r_h^2-1)^{3/2}+\pi^3 {Q_m}^4 (\beta -\alpha)}{8 \sqrt[4]{2} \sqrt{\pi} (e^{-\pi r_h^2}+\pi r_h^2 
-1)^{9/4}}.
\end{eqnarray}
\end{widetext}
The vector component of the vector field $\Psi=\left(\Psi^{r_h}, \Psi^\theta\right)$ yields
\begin{eqnarray}\nonumber
\Psi^{r_{h}}&=&\left(\frac{\partial \Psi}{\partial r_{h}}\right)_{{ Q_m, \theta}}\\ \nonumber
&=&\frac{\sqrt{\pi } e^{\pi r_h^2} (e^{\pi r_h^2}-1) r_h \csc (\theta) A}{16 \sqrt[4]{2} \sqrt[4]{C} [e^{\pi r_h^2} (\pi r_h^2-1)+1]^3}, \\ \label{2e}
\end{eqnarray}
where $A$ and $C$ are defined in Appendix A.
We discuss how ingoing and outgoing flows correspond to stable and unstable thermodynamic behaviour, respectively. The topological charge, obtained from the BH’s free energy, acts as a stability index. 

The normalized vector components are
\begin{widetext}
\begin{eqnarray}\label{2g} 
\frac{\Psi^{r_h}}{\|\Psi\|}&=&\frac{\pi A e^{\pi r_h^2} (e^{\pi r_h^2}-1) r_h \csc (\theta )}{\sqrt[4]{C} [e^{\pi r_h^2} (\pi r_h^2-1)+1]^3 (\csc^2 (\theta) \{B+4 \cot^2(\theta) [\sqrt{2} C^{3/2}-\pi C {Q_m}^2+\pi^3 {Q_m}^4 (\beta -\alpha)]^2\}/C^{9/2})^{1/2}},
\end{eqnarray}
\end{widetext}
where $A$, $B$ and $C$ are defined in {Appendix A}, and 
\begin{widetext}
\begin{equation}\label{2h}
\frac{\Psi^\theta}{\|\Psi\|}=-\frac{2 \cot (\theta)  [\pi ^3 {Q_m}^4 (\beta -\alpha )
-\pi  {Q_m}^2 C+\sqrt{2} C^{3/2}]}{ \{{4 \cot ^2(\theta ) [\pi ^3 {Q_m}^4 (\beta -\alpha )-\pi  {Q_m}^2 C+\sqrt{2} C^{3/2}]^2+B}\}^{1/2}}. 
\end{equation}
\end{widetext}
A normalized vector $n=\left(\frac{\Psi^{r_h}}{\|\Psi\|}, \frac{\Psi^\theta}{\|\Psi\|}\right)$ appeared in  Eqs.~$\left(\ref{2g}\right)$ and $\left(\ref{2h}\right)$.
In Fig.~\ref{fig2}, we depict the unit vector field in the $(r_h, \theta)$ plane for a BH solution in Einstein-Euler-Heisenberg theory. These unit vectors enable topological analysis of the system's critical points, where $r_0$ represents the characteristic length scale determined by the BH's event horizon geometry. 
\begin{figure}
\includegraphics[width=16pc]{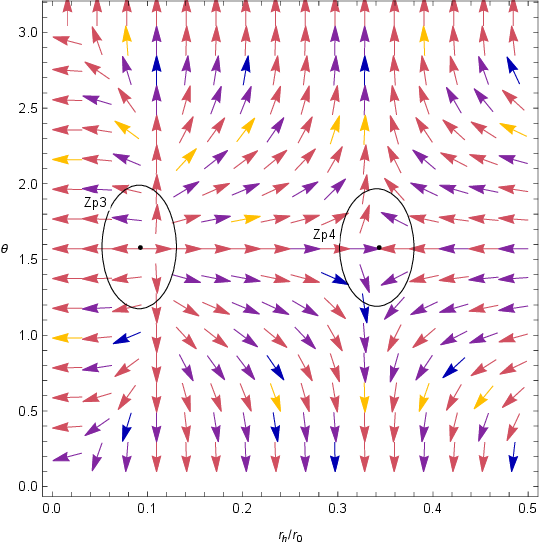}
\caption{\label{fig3} The field of normalized vector with fixed values of $\alpha =1.1$, $Q_m=1$, $\tau =30$ and $\Psi_m=1$.}
\end{figure}
The critical points are identified as $Z_{p1}$ (ingoing flow) and $Z_{p2}$ (outgoing flow), corresponding to stable and unstable regions of the system, respectively.
The ingoing flow ($Z_{p1}$) characterizes stable equilibrium configurations, while the outgoing flow ($Z_{p2}$) marks regions of instability in the field dynamics. We have computed the critical points by utilizing as $\theta=\frac{\pi}{2}$ in Eq.~$\left(\ref{2g}\right)$. The following illustrates a contour $C$ with parameters $\vartheta \in (0, 2 \pi)$ that can be utilized to obtain the topological charge through critical points  as \cite{i12,i13}
\begin{eqnarray}\label{o1}
r_{h}=  r_{0}+a \cos\vartheta, \hspace{.5cm}
\theta= \frac{\pi}{2}+b \sin\vartheta .
\end{eqnarray}
We study the details of these specific parameters as $(a, b, r_{0}) = (0.60, 0.20, 2\sqrt{3})$ and $(0.60,0.40,5.00)$. Here, red and blue curves show the $C_{1}$ and $C_{2}$ contours are shown in Appendix A. The deflection of the vector field $n$ is given by
\begin{eqnarray}\label{o2}
 \Omega (\vartheta) =\int^{\vartheta}_{0} \epsilon_{ab} n^{a}\partial_{\vartheta} n^{b} d\vartheta.
\end{eqnarray}
The topological charge is $Q = \frac{1}{2\pi}\Omega (2\pi)$, and this delimited by the contour $C_{1}$ and $C_{2}$, both are extracted to be zero. Thus, the total topological charge is $Q = 0$ in both contours. The function $\Omega (\vartheta)$ for $C_{1}$ and $C_{2}$ closer to $0$ at $\vartheta= 2\pi$.
\section{Defects of thermodynamic topology in canonical ensemble}\label{Sec:C}
At the moment, we are investigating the BH solution as topological thermodynamic defects in the canonical ensemble. The generalized free energy can be obtained using Eqs.~$\left(\ref{i2}\right)$-$\left(\ref{1}\right)$ as
\begin{eqnarray}  \nonumber
F=-\frac{e^{-\pi r_h^2}+\pi  r_h^2}{\tau }+\frac{{Q_m}^4 (\alpha -\beta)}{5 r_h^5}+\frac{{Q_m}^2}{2 r_h}+\frac{r_h}{2}.\\
\label{12}
\end{eqnarray}
The vector field components of the vector in Eq.~$\left(\ref{i4}\right)$ are given as follows
\begin{eqnarray}\nonumber
\Psi^{r_h}&=&\frac{2 \pi r_h (e^{-\pi r_h^2}-1)}{\tau}-\frac{{Q_m}^4 (\alpha -\beta)}{r_h^6}\\  \label{13}
&-&\frac{{Q_m}^2}{2 r_h^2}+\frac{1}{2},
\end{eqnarray}
where $C$ is defined in Appendix A and this can explain angular changes in spherically symmetric fields, as those found in electromagnetism/quantum mechanics. 
The vector field dependence on the polar angle is reflected in the trigonometric functions. Hence, the second component is represented as
\begin{equation}\label{14}
  \Psi^{\theta} =-\cot (\theta) \csc (\theta).  
\end{equation}
 The corresponding unit vectors are expressed as 
\begin{eqnarray}\label{15}
n^1=\frac{\frac{2 \pi r_h (e^{-\pi r_h^2}-1)}{\tau }+\frac{{Q_m}^4 (\beta -\alpha )}{r_h^6}-\frac{{Q_m}^2}{2 r_h^2}+\frac{1}{2}}{\left\{\left[H_1+\frac{{Q_m}^4 (\alpha -\beta )}{r_h^6}\right]^2+\cot^2(\theta) \csc ^2(\theta)\right\}^{1/2}},
\end{eqnarray}
and the second component as 
\begin{eqnarray}\label{16}       
n^2=-\frac{\cot (\theta) \csc (\theta)}{\left\{\left[H_1+\frac{{Q_m}^4 (\alpha -\beta)}{r_h^6}\right]^2+\cot ^2(\theta) \csc ^2(\theta)\right\}^{1/2}},
\end{eqnarray}
where
\begin{eqnarray}\label{17} 
H_1=-\frac{2 \pi  r_h (e^{-\pi r_h^2}-1)}{\tau }+\frac{{Q_m}^2}{2 r_h^2}-\frac{1}{2}.
\end{eqnarray}
By setting $\Psi^{r_h}=0$ and associating to zero points, the analytic expression for $\tau$ can be determine as
\begin{eqnarray}\label{18}
\tau=\frac{4 \pi  e^{-\pi r_h^2} r_h^7 (e^{-\pi r_h^2}-1)}{r_h^6-2 {Q_m}^4 (\alpha -\beta)-{Q_m}^2 r_h^4}.
\end{eqnarray}
With the help of critical points, we plot the unit vectors setting $\theta = \frac{\pi}{2}$ in Eq.~$\left(\ref{15}\right)$. We depict $n$ versus $\theta$ plane for the BH thermodynamic defects in Appendix A. Here, we observe that the three critical points named $CP_{4}$, $CP_{5}$ and $CP_{6}$ at $(a,b,r_{0})$= $(0.26,0.18,0.65)$, $(0.26,0.18,0.81)$ and $(0.27,0.18,1.71)$, respectively.
The contour $C_{3}$ and $C_{4}$ meet at $2 \pi$, we find that this has zero topological charge. We evaluate the stability of BH, it is simple to prove that BH with a large radius is stable while BH with a short radius is unstable, which is presented in Appendix A.
\section{Thermodynamic topology in mixed ensemble}\label{Sec:D}
In the present scenario, the magnetic potential $\Psi_m$ will be constant. It is expressed as 
\begin{eqnarray} \label{me1}
\Psi_m=\frac{Q_m}{r_h}.
\end{eqnarray}
The mass parameter can be derived as 
\begin{eqnarray} \label{me2}
M=\frac{2 {Q_m}^4 (\alpha -\beta)+5 {Q_m}^2 r_h^4+5 r_h^6}{10 r_h^5}.
\end{eqnarray}
Accordingly, a first-order phase transition can be examined through traditional critical point and find topological charge $-1$.
However, the emergence of a new critical point associated with a topological charge of $+1$  does not necessarily imply that a first-order phase transition encodes information about the ingoing flow (stable region of the system).
The modified temperature can be represented as
\begin{widetext}
\begin{eqnarray}  \label{me3}
T&=&\frac{-\pi  r_h^2 \Psi_m^2 (e^{-\pi r_h^2}+\pi  r_h^2-1)+\sqrt{2} 
(e^{-\pi r_h^2}+\pi r_h^2-1)^{3/2}+\pi ^3 r_h^4\Psi_m^4 (\beta -\alpha )}{8 \sqrt[4]{2} 
\sqrt{\pi }(e^{-\pi r_h^2}+\pi  r_h^2-1)^{9/4}}.
\end{eqnarray}
\end{widetext}
Thermodynamics function $\Psi$ is given as
\begin{widetext}
\begin{eqnarray}\label{me4}
\Psi &=& \frac{-\pi r_h^2 \Psi_m^2  (e^{-\pi r_h^2}+\pi  r_h^2-1)+\sqrt{2} (e^{-\pi r_h^2}+\pi  r_h^2-1)^{3/2}+\pi ^3 r_h^4 \Psi_m^4 (\beta -\alpha)}{8 \sqrt[4]{2} \sqrt{\pi} (e^{-\pi r_h^2}+\pi  r_h^2-1)^{9/4} \sin (\theta)}.
\end{eqnarray}
\end{widetext}
From the vector field $\Psi=\left(\Psi^{r_h}, \Psi^\theta\right)$, here we find the following
\begin{eqnarray}  \label{me7}
\frac{\Psi^{r_h}}{\|\Psi\|}=\frac{\sqrt{\pi } r_h \csc (\theta ) K}{16 \sqrt[4]{2} C^{3/4}  \left[\frac{\pi r_h^2 \csc^2(\theta ) K^2}{256 \sqrt{2} C^{13/2}}+H_2\right]^{1/2}},
\end{eqnarray}
where $H_2$ and $C$ are specified in Appendix A, and
\begin{eqnarray} \nonumber
\frac{\Psi^\theta}{\|\Psi\|}= -\frac{\cot (\theta) \csc (\theta) (l_1-\pi  r_h^2 C {\Psi_m}^2+\sqrt{2} C^{3/2})}{8 \sqrt[4]{2} \sqrt{\pi} C^{9/4} \left[\frac{\pi r_h^2 \csc ^2(\theta) K^2}{256 \sqrt{2} C^{13/2}}+H_{2}\right]^{1/2}},\\
\label{me8}
\end{eqnarray}
with
\begin{eqnarray} \label{meA8}
l_1=\pi ^3 r_h^4 {\Psi_m}^4 (\beta -\alpha).
\end{eqnarray}
These unit vectors can be used to examine the zero points. Here, we take the fixed values of $\alpha =0.1$ and $\Psi_m =1$. In this case, $r_{0}$ represents the arbitrary length scale based on the BH size surrounding the cavity. 
\section{ Defects of thermodynamic topology in mixed ensemble}\label{Sec:E}
In this section, we investigate the BH in Einstein Euler-Heisenberg theory as a topological defect in mixed ensemble. Nevertheless, we start with the potential for generalized free energy, given by 
\begin{eqnarray}\label{27}
F=M-Q_{m} {\Psi_m}-\frac{S_E}{\tau}.
\end{eqnarray}
In this case, from Eq.~$\left(\ref{27}\right)$, the modified mass parameter can be expressed as
\begin{eqnarray}\label{28}
M=\frac{{\Psi_m}^4 (\alpha -\beta)}{5 r_h}+\frac{1}{2} r_h ({\Psi_m}^2+1),
\end{eqnarray}
and
\begin{widetext}
\begin{eqnarray} \label{28a}
T=\frac{-\pi r_h^2 \Psi_m^2 (e^{-\pi r_h^2}+\pi r_h^2-1)+\sqrt{2} (e^{-\pi r_h^2}+\pi  r_h^2-1)^{3/2}+\pi ^3 r_h^4 {\Psi_m}^4 (\beta -\alpha)}{8 \sqrt[4]{2} \sqrt{\pi }(e^{-\pi r_h^2}+\pi r_h^2-1)^{9/4}}.
\end{eqnarray}
\end{widetext}
Moreover, generalized free energy potential can be achieved as
\begin{eqnarray}\nonumber
F=-\frac{e^{-\pi r_h^2}}{\tau}-\frac{\pi r_h^2}{\tau }+\frac{{\Psi_m}^4 (\alpha -\beta)}{5 r_h}-\frac{r_h}{2} ({\Psi_m}^2-1).
\\ \label{29}
\end{eqnarray}
This can be stated in terms of potential as
\begin{eqnarray}\nonumber
\Psi^{{r_h}}&=&\frac{-{2 {\Psi_m}^4 (\alpha -\beta )}/{r_h^2}-5 {\Psi_m}^2+5}{10} \\ \label{30} 
&+& \frac{2 \pi r_h (e^{-\pi r_h^2}-1)}{\tau},
\end{eqnarray}
and 
\begin{eqnarray}\label{31}
\Psi^{\theta}=-  \csc (\theta) \cot (\theta).
\end{eqnarray}
In this scenario, the intersection points are concurrent for greater $\tau$. 
The critical points associated with the destruction are easy to identify. The winding numbers of the two zeros are $\omega_1 = -1$ and $\omega_2 = 1$. Thus, the total topological number for the exact BH in Einstein Euler-Heisenberg is $\omega = \omega_1 + \omega_2 = 0$. The appropriate unit vectors are
\begin{eqnarray}\nonumber
n^{1}= \frac{H_3+{2 \pi r_h (e^{-\pi  r_h^2}-1)}/{\tau}}{\{{(\Psi^{\theta})^2+[H_3+2 \pi r_h (e^{-\pi r_h^2}-1) /{\tau}]^2}\}^{1/2}}, \\
\label{32}
\end{eqnarray}
where
\begin{eqnarray} \label{A32}
H_3= [-{2 {\Psi_m}^4 (\alpha -\beta)}/{r_h^2}-5 {\Psi_m}^2+5]/10,
\end{eqnarray}
and
\begin{eqnarray}\nonumber
n^{2}=\frac{\Psi^{\theta}}{\{(\Psi^{\theta})^2+[H_3+{2 \pi r_h (e^{-\pi r_h^2}-1) }/{\tau }]^2\}^{1/2}}. \\ 
\label{33}
\end{eqnarray}
By setting $\Psi^{r_h}=0$, one can find the analytical expression as
\begin{eqnarray}\label{34}
\tau=-\frac{20 \pi e^{-\pi r_h^2} (e^{\pi r_h^2}-1) r_h^3}{2 {\Psi_m}^4 (\alpha -\beta)+5 r_h^2 ({\Psi_m}^2-1)}.
\end{eqnarray}
In Fig.~\ref{fig3}, we show the normalized vector in a mixed ensemble and identify the critical points. The critical points are identified as $Z_{p3}$ (outgoing flow) and $Z_{p4}$ (ingoing flow), corresponding to stable and unstable regions at $(\theta,r_h/r_{0})= (1.50, 0.10)$, and (1.50, 0.35), respectively. The outgoing flow ($Z_{p3}$) characterizes stable equilibrium configurations, while the ingoing flow ($Z_{p4}$) marks regions of instability in this case. The contour $C_{1}$ (red curve) first increases with certain values and then decreases to meet with the $\theta$-axis at $2 \pi$, the detail is discussed in Appendix A. So far, after a loop, $\Omega$ vanishes, which implies that the charge is zero. The critical point for $\frac{\tau}{r_{0}}=30$ is located at $(a,b,r_{0})=(0.50, 1.23,2.61)$.
\section{Thermal geometries with exponential framework}\label{Sec:F}
\begin{figure}
 \includegraphics[width=14pc]{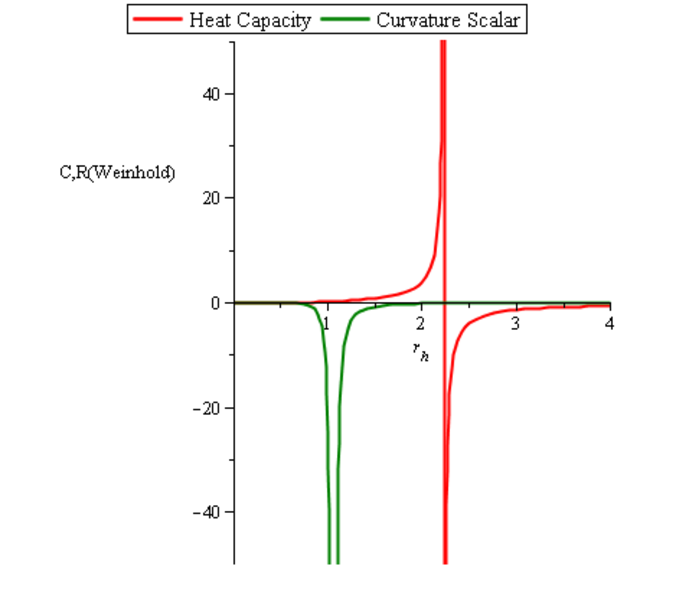} 
 \caption{\label{fw}  Weinhold curvature scalar (green curve) $R^{\mathrm{W}}$ and heat capacity (red curve) with fixed values of $\alpha =3$, $\beta =2.1$ and ${Q_m}=1.5$.}
\end{figure}

In Einstein Euler-Heisenberg theory, we explore the geometrical structures of the precise BH through Ruppiner, Weinhold, GTD and HPEM curvature scalars. On the other hand, we articulated the positive (stable) BH behavior for four specific horizon radius ranges which provide the attracting and repulsive behavior of the BHs. We will compare our results with existing approaches from the literature. Further details and a comprehensive discussion can be found in Refs.~\cite{g3,g4,g8,g9, Aman:2003ug}. There exists a connection between the divergence and critical points associated with the scalar curvature. The representative line element of the BHs is described by
\begin{equation}\label{s1}
d s^2=\frac{S_E M_{S_E}}{\left(\prod_{i=2}^n \frac{\partial^2 M}{\partial \xi_i^2}\right)^3}\left[\Sigma_{i=2}^n\left(\frac{\partial^2 M}{\partial \xi_i^2}\right) d \xi_i^2-M_{S_E S_E} d S_E^2\right].
\end{equation}
Here, $M_S$ represents the derivative of mass $M$ with respect to entropy $S_E$ and $\xi_i \neq S_E$. The intrinsic expression of the metric space is as follows
\begin{equation}\label{s2}
g=\left(E^c \frac{\partial  \Pi}{\partial E^c}\right)\left(\eta_{a b} \delta^{b c} \frac{\partial^2 \Pi}{\partial E^c \partial E^d} d E^a d E^d\right).
\end{equation}
The extensive, intensive, and thermodynamic potential parameters are denoted by $\frac{\partial \Pi}{\partial E^c}=\delta_{b c} I^b$, $\Pi$, $E^a$, and $I^b$, respectively. Thermodynamic geometries including Ruppeiner, Weinhold,  HPEM, and GTD formalisms are covered in this section. The theory of BH thermodynamics has been extensively studied in the literature, but there hasn't been much discussion of BH microstructure. Weinhold geometry is illustrated as
\begin{equation}\label{s3}
g^{\mathrm{W}}_{i,j}=\partial_i \partial_j M(S_E,Q_m).
\end{equation}
The Weinhold metric is given by
\begin{equation}\label{s4}
ds^2_\mathrm{W}=M_{S_E S_E}dS_E^2+M_{Q_m Q_m}dQ_m^2+2 M_{S_E Q_m}dS_E dQ_m,
\end{equation}
with the following matrix form
\begin{equation}\label{s5}
\begin{pmatrix}
M_{S_E S_E} & M_{S_E Q_m}\\ 
M_{Q_m S_E} & M_{Q_m Q_m} \\
\end{pmatrix}.
\end{equation}
The curvature scalar $R^{\mathrm{W}}$ can be computed as follows using the above described equations as
\begin{eqnarray} \nonumber
R^{\mathrm{W}}&=&({30}/{7}) [-({15 \sqrt {2}}/{7}) w_1 (-1+S_E )^{7/2}+ w_2 {Q_m}^{4} {\pi}^{3}  \\ \nonumber
&&(-1+ S_E) ^{3} (\beta-\alpha) ] \sqrt [4]{2}\sqrt{\pi} (-1+S_E)^{-{3/4}}  w_3^{-2}.
\\
\label{As5} 
\end{eqnarray} 
We illustrate both the positive and negative behavior of $R^{\mathrm{W}}$ in this case. Similarly, we compare the behavior of the curvature (green curve) with that of the heat capacity (red curve), as shown in Fig.~\ref{fw}. On the other hand, we noticed the negative behavior after the singular point such as $r_h=2.25$. 

Furthermore, we studied the attractive behavior in Einstein Euler-Heisenberg theory, uncovering key phenomenological details of molecular attraction. 
The scalar curvature $R^{\mathrm{W}}$ of the Ruppeiner geometry which is conformally equivalent to the Weinhold geometry can be determined as follows
\begin{equation}\label{s7}
ds^{2}_{R}=\frac{ds^{2}_{W}}{T}.
\end{equation}
The explicit expression of Ruppeiner geometry is calculated as
\begin{eqnarray} \nonumber
R^{\mathrm{Rup}}&=&({240 \sqrt {2}}/7) (-1+S_E)^{3/2}\pi x_1 [(9 \sqrt {2}/7) ({Q_m}^
{2} (-\alpha \\  \nonumber
&+&\beta) {\pi }^{2}+5/6-5 S_E/6) (-1+S_E)^{3/2}+ x_2 \pi\\  \label{RUP1} 
 &&{Q_m}^{2}]^{-2} (\sqrt {2} (-1+S_E)^{3/2}+
 x_3\pi {Q_m}^{2})^{-1}.
\end{eqnarray}
Here, $x_1$, $x_2$ and $x_3$ are presented in Appendix A. Thus, since the curvature scalar  (green curve), we observed the thermodynamic phase change of the Ruppiner metric in terms of $S_E$. A new metric HPEM was presented in development of geometric phase space by utilizing the thermodynamic quantities. 
We demonstrate the Ricci scalar $R^{\mathrm{HPEM}}$ behavioral characteristics in the metric, this allows us to identify the type of phase transition and curvature can be measured by the thermodynamic interaction. The HPEM geometry is represented as
\begin{equation} 
ds^2=\frac{S_E M_{S_E}}{\left(\frac{\partial^2 M}{\partial Q_m^2}\right)^3}(-M_{S_E S_E}dS_E^2+M_{Q_m Q_m}dQ_m^2). \\ 
\label{s9}
\end{equation}
The mathematical expression for the HPEM geometry scalar is calculated as
\begin{eqnarray} \nonumber
R^{\mathrm{HPEM}}&=&({{4375}/{648}}) S_E y_1 {2}^{3/4}  (-1+S_E)^{3/4}  y_2\\  \nonumber
&&{\pi }^{1/2}[-5+5 S_E+6 {\pi}^{2}{Q_m}^{2} (\alpha-\beta)]^{-3}\\ \nonumber
&&[(\sqrt {2}/3) (-1+S_E)^{3/2}+\pi  y_3 {Q_m}^{2}]^{-2} \\ \label{s10}
&&[ {Q_m}^{2} (\beta-\alpha) {\pi }^{2}+5/6-5S_E/6]^{-2}.
\end{eqnarray}
Here, $y_1$, $y_2$ and $y_3$ are presented in  Appendix B. The divergence of the HPEM metric's scalar curvature  (green curve) and heat capacity  (red curve) at zero point is observed. As a result, we derive some useful information from HPEM framework. The GTD metric's scalar curvature can be expressed as \begin{equation}\label{s11}
ds^2=(-S_E M_{S_E S_E}+ Q_m M_{Q_m Q_m }) \begin{pmatrix}
M_{S_E S_E} & 0\\
0 & M_{Q_m Q_m}\\
\end{pmatrix}.
\end{equation}
From the above expression, one can obtain 
\begin{eqnarray} \nonumber
R^{\mathrm{GTD}}&=& ({35J_1}/{288}) \{(\sqrt {2} S_E/3) (-1+{S_E})^{3/2}\\   \nonumber
&+& [(Q_m S_E-({64}/{15}) (-1+S_E)^{2}) (\beta-\alpha) {Q_m}^{2}\\   \nonumber
&&{\pi }^{2}-5 Q_m S_E/9 -(32/5) (-1+S_E)^{2}](-1\\\nonumber
&+&S_E)\} \pi Q_m) (-1+S_E)^{-4}[{Q_m}^{2} (\beta-\alpha) {\pi}^{2}\\  \label{s11}  
&+&5/6-(5/6) S_E]^{-2} (J_2)^{-2}.
\end{eqnarray}
Here, $J_1$ and $J_2$ are defined in Appendix B. In Fig.~\ref{fgtd}, we depict the scalar behavior (green curve) of the GTD curvature. The distinctive point for the curvature scalar at $r_h = 1$. In this case, the heat capacity is not coincident (meets) with the unique point $r_h = 1$. 
Therefore, the GTD geometry does not provide us with any physical information about this framework.
\begin{figure}
 \includegraphics[width=14pc]{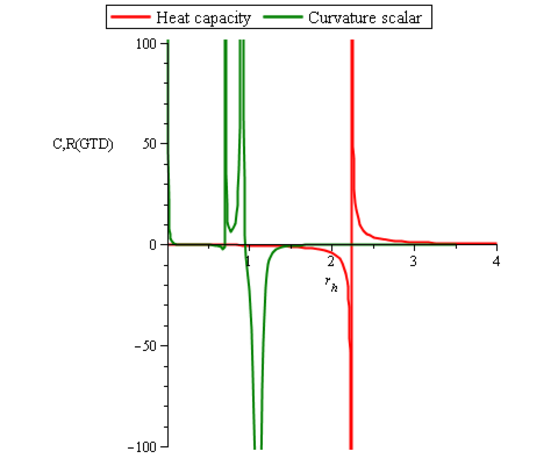}
 \caption{\label{fgtd} GTD curvature scalar (green curve) $R^{\mathrm{GTD}}$ and heat capacity (red curve) with fixed values of $\alpha =3$, $\beta =2.1$ and ${Q_m}=1.5$.}
\end{figure} 
\begin{table}[h!]
\centering
\caption{Theoretical comparison of thermodynamic geometries}
\begin{tabular}{l c c c c c}
\hline
\hline
{Method}  & {Heat capacity variation} &  {R,C behavior} \\ 
\hline
Weinhold & $1.40<r_h<3.50$ & coincidence: $r_h=1.75$\\ 
Ruppeiner & $1.80<r_h<2.80$ & coincidence:  $r_h=1.60$  \\ 
HPEM &   $2.00<r_h<3.00$ & coincidence:  $r_h=1.10$  \\ 
GTD  &  $1.50<r_h<2.80$ & coincidence:  $r_h=1.50$  \\ 
\hline 
\hline
\end{tabular}
\label{table:medium}
\end{table}
In Table \ref{table:medium}, we summarize the attractive (negative) and repulsive (positive) nature of thermodynamic curvature, On the other hand, no interaction between particles is shown by a curvature of zero with values of $\alpha = 3$, $\beta = 2.1$, and $Q_m = 1.5$.

\section{ Emission energy with exponential framework}\label{Sec:G}

It is well-known established that, in the immediate vicinity of a BH horizon, quantum fluctuations generate an excess of particles, which are subsequently annihilated within interior of BH. 
In the core region where Hawking radiation occurs, positive-energy particles tunnel out of the BH, which is the main source of BH evaporation during the process.
We intend to have a conversation about the energy emission rate in this part. At an extremely high phase, the BH's energy reception  in  cross-section oscillates until it reaches to the  limiting and constant value $\sigma_\mathrm{lim}$ as \cite{EE1}
\begin{eqnarray}\label{EE1}
\sigma_\mathrm{lim} \approx \pi R_0^2.
\end{eqnarray}
The BH horizon radius is denoted by $R_0$. The energy emission rate expression for BH is given by \cite{g9, EE1}
\begin{eqnarray}\label{EE2}
\frac{d^2 \varepsilon}{d \omega dt}=\frac{2 \pi^2 \sigma_\mathrm{lim}}{\exp (\frac{\omega}{T})-1} \omega^3,
\end{eqnarray}
where the Hawking temperature is expressed  in  Eq.~$\left(\ref{2d}\right)$. According to this,  we drive the Eq.~$\left(\ref{EE2}\right)$ after the replacement of horizon radius $r_0$, temperature $T$  and cross-section $\sigma_\mathrm{lim}$ as
\begin{eqnarray}\label{ee3}
\frac{d^2 \varepsilon}{d \omega d t}= \frac{2 \pi ^3 r_h^2 \omega ^3}{\exp \left[\frac{8 \sqrt[4]{2} \sqrt{\pi } \omega  (e^{-\pi r_h^2}+\pi  r_h^2-1)^{9/4}}{v_1}\right]-1}.
\end{eqnarray}
Here, ${v_1}$ is presented in Appendix B. In Fig.~\ref{fee2}, we show how the rate of energy emission varies for different values of $\beta$.
\begin{figure}
 \includegraphics[width=16pc]{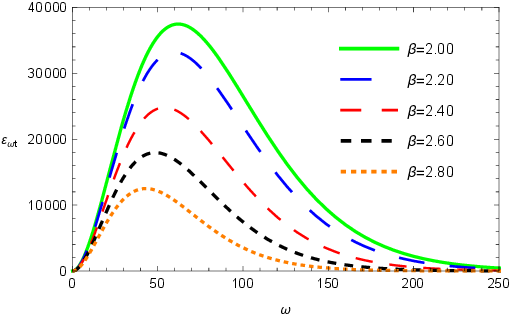}
 \caption{\label{fee2} Rate of energy emission with fixed values of $\alpha =0.6$, ${Q_m}=0.1$, $\beta$ = 2.00 (green curve), 2.20 (blue dashed curve),  2.40 (red dashed curve),  2.60 (black dashed curve), and  2.80 (orange dashed curve).}
\end{figure}
Our study reveals that it increases the parameter for $\beta$ = 2.00 (green curve), 2.20 (blue dashed curve), 2.40 (red dashed curve), 2.60 (black dashed curve), and  2.80 (orange dashed curve), which in turn significantly enhances the energy emission rate. We systematically analyze how charged BH emit energy differently as their $\beta$ increases, showing numerous characteristics phases where quantum effects strongly suppress Hawking radiations.

\section{Conclusions} \label{Sec:L}

In this paper, we have computed the thermodynamic quantities with exponential entropy of BH in Einstein Euler-Heisenberg theory. The thermodynamics, stability, and divergence for BH are then thoroughly examined.  We have studied the critical and divergence points based on coupling factors and the thermal stability of BH.  For canonical, mixed, and grand canonical ensemble scenarios, we looked into cases where the total topological charge is either $0$ or $-1$.  However, in mixed, and grand canonical ensembles that have $0$ topological charge.    

In the first part, we observed that a canonical ensemble consists of magnetic charge through magnetic potential and a grand canonical ensemble which maintained the consistency through magnetic potential. We examined the novel as well as the conventional topological charges of each ensemble through critical points. Subsequently, while conducting an in-depth study, we obtained that the topological charge $Q = -1$. It is expected that the new topological argument would clarify the BH astronomical occurrences and offer an insightful concept for the investigation of BH photon spheres and  light rings. Additionally, we depicted thermodynamic variables in terms of the horizon's radius $r_h$. We demonstrated the single zero point via temperature and heat capacity for each maximum as well as the minimum value of the mass parameter. 

In the second part, we examined the zero points in Ruppeiner and the Weinhold metrics through thermodynamic geometry method in  Fig.~\ref{fw}. We investigated the features of different values of the spacetime parameters on the stability conditions of  BH in Einstein Euler-Heisenberg theory. It follows from the considerations in Refs. \cite{Aman:2003ug,R1,R2} that the singular points of HPEM geometry coincide with the heat capacity's zero points. Moreover, we studied another HPEM measure the divergence which corresponds to the singular points of heat capacity.

As a result, we gained deeper insights into black hole thermodynamics by comparing the HPEM geometry with the other metrics discussed in the referenced works \cite{Aman:2003ug,R3}. The GTD metric may explain the divergence point of the heat capacity for the static charged BH in Fig.~\ref{fgtd}. Lastly, Fig.~\ref{fee2} provides more detail on the energy emission rate which is reliant on the frequency $\omega$. 
We systematically analyze the energy emission of charged black holes as their $\beta$ parameter increases, revealing distinct characteristic phases in which quantum effects significantly suppress Hawking radiation \cite{R4}.

By using these geometric methods, we discover complex information on the statistical characteristics and thermodynamic interactions of the BH's minuscule degrees of freedom.  By examining how these techniques interact, we were able to spot new microstructure patterns like phase transitions and critical occurrences that shed light on the fundamental geometric and quantum characteristics of BHs.  The specific exponential entropy creates and positions these charges in the phase space.
Therefore, analyzing the sum and nature of these topological charges provides direct insight into the BH’s thermodynamic stability and identifies the phases where stability transitions occur, thereby linking topology to stability through the underlying structure of entropy \cite{Rani:2024qju}.  This approach uncovers novel microstructure signatures such as phase transitions and critical phenomena that elucidate fundamental geometric and quantum features of BHs.
These results not only enhance our understanding of BH thermodynamic topology but also pave the new way for broader features in modified theories of gravity and quantum gravity frameworks. Investigating exponential entropy effects on thermodynamic stability and phase transitions can reveal novel aspects of BH behavior in strong-field regimes.
\section*{Acknowledgement}
The work was sponsored by the National Natural Science Foundations of China (Grant Numbers: 12235007, 12375003). The work of KB was also supported by the JSPS KAKENHI Grant Numbers 21K03547, 24KF0100 and Competitive Research Funds for Fukushima University Faculty (25RK011).
\begin{figure}
\includegraphics[width=14pc]{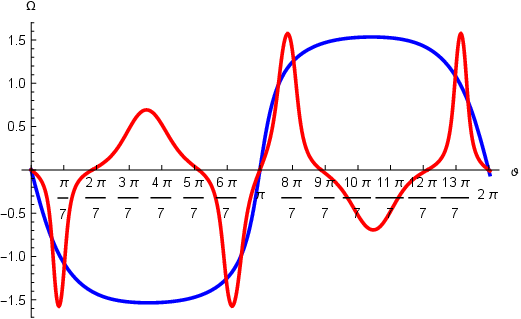}
\caption{\label{f5} $\Omega$ vs $\vartheta$ for the contours $C_1$ (red curve) and $C_2$ (blue curve) of fixed values  $a=1.8$,  $\alpha =0.3$,  $b=4.22$, $\beta =0.5$, ${Q_m}=5.5$, ${r_0}=3.7$, $\tau =5.5$ and  ${\Psi_m}=0.69$.}
\end{figure}
\section*{Appendix A: Thermodynamic topology }

Abbreviations and some long calculations are defined in this Appendix.
\begin{figure}
\includegraphics[width=15pc]{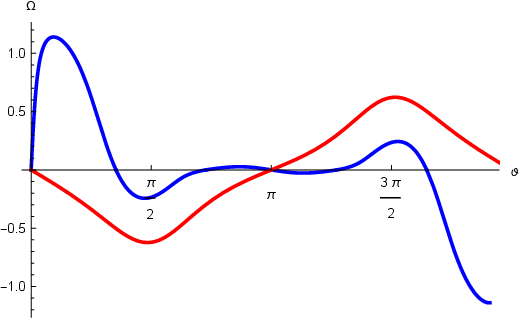}
\caption{\label{f7} $\Omega$ vs $\vartheta$ for the contours $C_1$ (red curve) and $C_2$ (blue curve) of fixed values $a=3.2$, $\alpha =0.6$, $b=0.52$, $\beta =0.2$, ${Q_m}=1.5$, ${r_0}=2.1$, $\tau =3.5$ and  $\Psi_m=0.69$.}
\end{figure}
In Fig.~\ref{f7}, we depict the contours  $C_{1}$ (red curve)  and $C_2$ (blue curve), which exhibit non-monotonic behavior: both contours initially rise, reach a maximum, and then decay asymptotically to intersect the  $\theta$-axis at $2 \pi$. This reveals the phase transition and stability via their shape, enclosed areas and  intersections. The vortex patterns mark stable ($+1$) and unstable ($-1$) quantified by winding numbers, respectively. Asymmetries curves provide the symmetry breaking boundaries. After a loop, $\Omega$ vanishes and this implies that the charge is zero.
In Fig.~\ref{f5}, we analyzed the two contours $C_{3}$ (red curve) and $C_{4}$ (green curve) for fixed values of $\alpha =2.1$ and $Q_m=1$. Here, red and blue curves are defined by $C_{3}$ and $C_{4}$ contours, respectively. The contours $C_{3}$  and  $C_{4}$ meet at $2 \pi$, it means that having zero topological charge. Our stability analysis shows that BH with small radii are unstable (exhibiting negative stability), while larger-radius configurations demonstrate stable behavior (with positive stability). The critical point for $\frac{\tau}{r_{0}}=30$ is located at $(a,b,r_{0})=(0.50, 1.22,2.60)$. The expression of $A$ is given by
\begin{align} \nonumber
A&= 9 \pi^3 {Q_m}^4 e^{\pi r_h^2} (\alpha -\beta)+5 \pi^2 {Q_m}^2 e^{\pi r_h^2} r_h^2 \\  \nonumber
&-5 \pi {Q_m}^2 (e^{\pi r_h^2}-1)-3 r_h^2 \sqrt{2} \pi e^{\pi r_h^2} \sqrt{C} \\ \label{A1} 
&+3 \sqrt{2 C} (e^{\pi r_h^2}-1). 
\end{align}
The constants play a key role in topology thermodynamics by preserving dimensional consistency, improving complex expressions for easier comprehension, and establishing a direct connection between topological attributes and physical thermodynamic parameters.
We find that the expression of $B$ is described as
\begin{equation}\label{A2}
B=\frac{\pi^2 A^2 e^{- \pi r_h^2} (e^{\pi r_h^2}-1)^2 r_h^2}{[e^{\pi r_h^2} (\pi  r_h^2-1)+1]^2}, 
\end{equation}
and the expression of $C$ is given by
\begin{align}\label{A3} 
C=\pi r_h^2+e^{-\pi r_h^2}-1.
\end{align}
Moreover, the expression of $H_2$ can be formulated as
\begin{widetext}
\begin{align}\label{A4}
H_2=\frac{\cot^2(\theta) \csc^2(\theta) [\pi ^3 r_h^4 {\Psi_m}^4 (\beta -\alpha)-\pi r_h^2 C {\Psi_m}^2+\sqrt{2} C^{3/2}]^2}{64 \sqrt{2} \pi C^{9/2}}.
\end{align}   
\end{widetext}
This ensures that the outcomes are comprehensible and related to the underlying physical system frameworks
and 
\begin{align}\nonumber
K&=2 C [4 \pi ^2 r_h^2 {\Psi_m}^4 (\beta -\alpha)-2 \pi r_h^2 {\Psi_m}^2 (1-e^{-\pi  r_h^2})\\ \nonumber
 &- 2C {\Psi_m}^2+3 \sqrt{2 C} (1-e^{-\pi r_h^2})]-9 (1-e^{-\pi r_h^2})\\ \label{A5}
&[\pi^3 r_h^4 {\Psi_m}^4 (\beta -\alpha)-\pi  r_h^2 C {\Psi_m}^2+\sqrt{2} C^{3/2}]. 
\end{align}
\section*{Appendix B: Thermal Geometries}
In this Appendix, Weinhold geometry employs normalized variables to ensure dimensionless geometric characteristics, enhancing interpretability while preserving dimensional consistency in the geometric methods. In order to emphasize important thermodynamic features, algebraic forms should be simplified.
\begin{figure}
 \includegraphics[width=15pc]{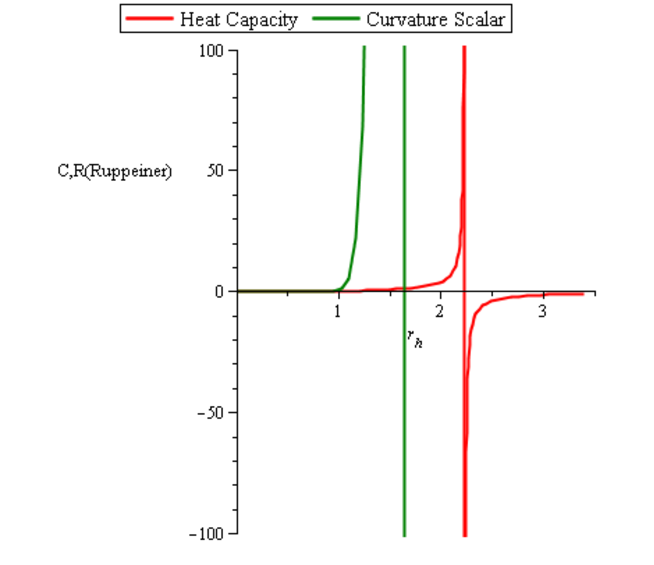}
 \caption{\label{frup} Ruppeiner curvature scalar (green curve) $R^{\mathrm{Rup}}$ and heat capacity (red curve) with fixed values of $\alpha =3$, $\beta =2.1$ and ${Q_m}=1.5$.}
\end{figure}
The negative values in this instance provide fictitious roots. The scalar curvature (green curve) coincides with the heat capacity (red curve) and temperature zero points, signifying the phase changeover point. In Fig.~\ref{frup}, we show the curvature scalar (green curve) $R^{\mathrm{Rup}}$ of BH concerning the horizon radius.
The resulting curvature scalar, derived from the above equation, is plotted against the horizon radius to investigate the phase transition by comparing it with the zeros of the heat capacity. Additionally, the Ruppeiner geometry can be analyzed for fixed parameters $\alpha$ and $\beta$, with the results corroborated by the plotted graph. 
\begin{figure}
 \includegraphics[width=14pc]{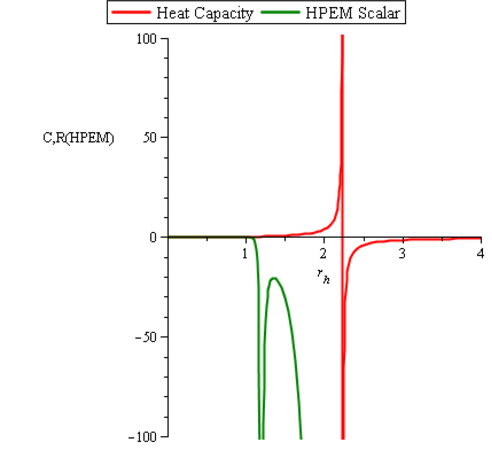} 
 \caption{\label{fhpem}  HPEM curvature scalar (green curve) $R^{\mathrm{W}}$ and heat capacity (red curve)  with fixed values of $\alpha =3$, $\beta =2.1$ and ${Q_m}=1.5$.}
\end{figure}
 In Fig.~\ref{fhpem}, we explore the divergence points of the HPEM metric's Ricci scalar which provide the two different kinds of heat capacity phase transitions. 
The representation of $w_1$ is as follows
\begin{align} \nonumber
w_1&={Q_m}^{4} {\pi}^{4}(\beta-\alpha)^{2}-(7 {\pi }^{2} {{Q_m}}^{2}/5) (\beta-\alpha)  (-1 \\ \label{B1}
&+S_E)+(-1+S_E )^{2}/4,
\end{align} 
and the form of $w_2$ is given by
\begin{align}  \label{B2}
 w_2={Q_m}^{2} {\pi}^{2} (\beta-\alpha)-25 S_E/{42}+{25}/{42}.
\end{align}
Additionally, geometric quantities should be connected to physically significant properties like energy and temperature. Moreover, maintains the thermodynamic system's symmetries and making it easier to compare the results to existing findings or other geometric frameworks in thermodynamics. The expression of $w_3$ is represented as
\begin{align} \nonumber
w_3&=({9}/{7}) [{Q_m}^{2} {\pi}^{2} (\beta-\alpha)+5/6-5S_E/6] \sqrt{2} (-1\\  \nonumber
&+S_E)^{3/2}+\pi {Q_m}^{2}[{Q_m}^{4} {\pi}^{4}(\beta-\alpha) ^{2}-(5/2) {Q_m}^{2}
\\ \label{B3}
&{\pi}^{2}(\beta-\alpha) (-1+S_E) +(15/14) (-1+S_E)^{2}].
\end{align}
Certain constants can be added to Ruppeiner geometry, a particular type of thermodynamic geometry, in order to make the mathematical formulas involved simpler or less complex. The expression $x_1$ can be formulated as
\begin{align}  \nonumber
x_1&=-({15 \sqrt {2}}/{7})[{Q_m}^{4} (-\alpha+\beta)^{2}{\pi}^{4}-(7/5){Q_m}^{2}
(-\alpha \\  \nonumber
&+\beta)(-1+S_E) {\pi}^{2}+(1/4)(-1+S_E) ^{2}] (-1+S_E)^{7/2}\\  \nonumber
&+ [{Q_m}^{2} {\pi}^{2}(-\alpha+\beta) -{25 S_E}/{42}+{{25}/{42}}](-1\\ \label{B4}
&+S_E)^{3}{\pi}^{3}{Q_m}^{4} (-\alpha+\beta).
\end{align}
To keep the equations' dimensional consistency, we have defined some new parameters in temperature, pressure, and volume to guarantee the resultant expressions that have consistent units. Also, these values are important and have physical significance.  The expression of $x_2$ is represented as
\begin{align} \nonumber
x_2&={Q_m}^{4} {\pi }^{4} (-\alpha+\beta) ^{2}-(5 {\pi}^{2} {Q_m}^{2}/2) (-\alpha+\beta)(-1 \\ \label{B5}
&+S_E)+({15}/{14}) (-1+S_E)^{2}.
\end{align}
Occasionally, these constants are included to connect the observable thermodynamic qualities to abstract the geometric quantities, and we find that $x_3$ is expressed as
\begin{align} \label{B6}
x_3= {Q_m}^{2} {\pi}^{2} (-\alpha+\beta)-{S_E}+1.
\end{align}
Stabilizing variables to facilitate analysis and interpretation, and simplifying intricate statements to emphasize important thermodynamic behaviors, and $y_1$ displayed as follows
\begin{align} \nonumber
y_1&=\sqrt {2} (-1+S_E)^{3/2}-\pi {Q_m}^{2} (-1+S_E)\\ \label{B7}
&+{\pi }^{3}{Q_m}^{4} (\beta-\alpha). 
\end{align}
Additionally, we relate the geometric properties to quantifiable physical quantities to maintain the symmetries of the system which ensures the outcomes in thermodynamic frameworks. This makes it simpler to compare the results with other geometric approaches such as Ruppeiner and Weinhold geometry, and $y_2$ is given by 
\begin{align} \nonumber
y_2&=-({18 \sqrt {2}}/{7})[\beta-\alpha^{2}{\pi}^{4}-(7/6){Q_m}^{2}
(\beta-\alpha) (-1\\ \nonumber
&+S_E) {\pi }^{2}+({5}/{27}) (-1+S_E)^{2}] (-1+S_E)^{7/2}+{\pi}^{3} (-1\\  \nonumber
&+{S_E})^{3}{Q_m}^{4}(\beta-\alpha)[{Q_m}^{2} (\beta-\alpha) {\pi}^{2}-{{25 S_E}/{42}} \\ \label{B8}
&+ {{25}/{42}}],
\end{align}
and $y_3$ is expressed as
\begin{align} \label{B9}
y_3= {Q_m}^{2} {\pi}^{2} (\beta-\alpha)+5/9-5S_E/9. 
\end{align}
Also, some new parameters are used in GTD geometry to preserve the system in symmetries, and the comparisons with other thermodynamic frameworks to make it easier and dimensional consistency is assured.  The expression of $J_1$ is given by
\begin{align} \nonumber
J_1 &= -({18 \sqrt {2}}/{7}) [{{Q_m}}^{4} (\beta-\alpha)^{2}{\pi}^{4}-(7/6){Q_m}^{2}(\beta-\alpha) (-1\\ \nonumber
&+S_E) {\pi}^{2}+({5}/{27})(-1+S_E)^{2}] (-1+{S_E})^{7/2} +{\pi}^{3} (-1\\ \nonumber
&+S_E)^{3}{{Q_m}}^{4} (\beta-\alpha)[{Q_m}^{2}(\beta-\alpha) {\pi}^{2}-{{25 S_E}/{42}} \\ \label{B10}
&+{{25}/{42}}].
\end{align}
We simplify the complex expressions for easier analysis and maintain the relationship between geometric quantities and physical thermodynamic properties. The expression of $J_2$ is represented as
\begin{eqnarray}\nonumber
J_2&=&  \sqrt {2} (-1+S_E)^{3/2}/3+\pi{Q_m}^{2} [{Q_m}^{2} {\pi}^{2} (\beta-\alpha) +5/9\\  \label{B11}
&-&5S_E/9].
\end{eqnarray}
Here, $v_1$ is introduced to ensure dimensional consistency, simplify the expressions for easier computation, and relate the abstract mathematical terms directly to physical parameters like temperature, and frequency, making the results more interpretable and aligned with known physical laws.
We find that $v_1$ is expressed as
\begin{align}\nonumber
v_1&=-\pi  {Q_m}^2 (e^{-\pi r_h^2}+\pi  r_h^2-1)+\sqrt{2} e^{-\pi r_h^2}. \label{C1}
\end{align}

\end{document}